\documentclass[prb,aps,superscriptaddress,reprint]{revtex4-1}

\usepackage{graphicx}% Include figure files
\usepackage{dcolumn}% Align table columns on decimal point
\usepackage{bm}% bold math
\usepackage{setspace}
\usepackage{todonotes}

\newcommand{\fg}{f_{\mathrm{G}}}
\newcommand{\vc}{v_{\mathrm{crit}}}

\newcommand{\hset}{H_{\mathrm{set}}}
\newcommand{\hrf}{h_{\mathrm{rf}}}
\newcommand{\hip}{H_{\mathrm{IP}}}

\begin{document}

\title{Electrical measurement of magnetic-field-impeded polarity switching of a ferromagnetic vortex core}

\author{Manu~Sushruth}
\affiliation{School of Physics, M013, University of Western Australia, 35 Stirling Hwy, Crawley WA 6009, Australia.}%
\author{Jasper P.~Fried}
\affiliation{School of Physics, M013, University of Western Australia, 35 Stirling Hwy, Crawley WA 6009, Australia.}%
\author{Abdelmadjid~Anane}
\affiliation{Unit\'e Mixte de Physique, CNRS, Thales, Univ.~Paris-Sud, Universit\'e Paris-Saclay, 91767, Palaiseau, France.}%
\author{Stephane~Xavier}
\affiliation{Thales Research and Technology, 1 Avenue A.~Fresnel, Palaiseau, France}
\author{Cyrile~Deranlot}
\affiliation{Unit\'e Mixte de Physique, CNRS, Thales, Univ.~Paris-Sud, Universit\'e Paris-Saclay, 91767, Palaiseau, France.}%
\author{Mikhail~Kostylev}
\affiliation{School of Physics, M013, University of Western Australia, 35 Stirling Hwy, Crawley WA 6009, Australia.}%
\author{Vincent~Cros}
\affiliation{Unit\'e Mixte de Physique, CNRS, Thales, Univ.~Paris-Sud, Universit\'e Paris-Saclay, 91767, Palaiseau, France.}%
\author{Peter J.~Metaxas}
\email{peter.metaxas@uwa.edu.au}
\affiliation{School of Physics, M013, University of Western Australia, 35 Stirling Hwy, Crawley WA 6009, Australia.}%

\date{\today}% It is always \today, today,
             %  but any date may be explicitly specified

%%%%%%%%%%%%%%%% ABSTRACT %%%%%%%%%%%%%%%%

%150 words
\begin{abstract}
Vortex core polarity switching in NiFe disks has been evidenced using an all-electrical rectification scheme. Both simulation and experiments yield a consistent loss of the rectified signal when driving the core at high powers near its gyrotropic resonant frequency. The frequency range over which the loss occurs grows and shifts with increasing signal power, consistent with non-linear core dynamics and periodic switching of the  core polarity induced by the core attaining its critical velocity. We demonstrate that core polarity switching can be impeded by displacing the core towards the disk's edge where an increased  core stiffness reduces  the maximum attainable core velocity.
\end{abstract}

\maketitle

Ferromagnetic vortices are curled magnetization configurations with out of plane magnetized cores\cite{Cowburn1999-2,Shinjo2000,Wachowiak2002}. They arise naturally\cite{Guslienko2008} at zero external magnetic field in thin ferromagnetic discs with negligible intrinsic anisotropy and diameters $\sim 0.1 - 1$ $\mu$m. An example of a magnetic topological defect or soliton, ferromagnetic vortices have been the subject of much applications-driven research typically focused on data storage\cite{Yamada2007}, logic\cite{Jung2011} and radiofrequency nano-devices\cite{Pribiag2007,Dussaux2010}. 

Spintronic\cite{Pribiag2007,Dussaux2010} and magnonic\cite{Sugimoto2011,Huber2011}  applications often exploit the vortex core's gyrotropic resonance which involves a confinement-influenced gyration of the core around its equilibrium position.
The gyrotropic mode can be driven by spin transfer torques or in-plane oriented r.f.~magnetic fields.
At low excitation powers, this resonance entails a steady-state gyration of the vortex core at a well defined frequency\cite{Novosad2005}, $\fg$. However,  excitation at higher powers can generate more complex, but fundamentally appealing, non-linearities such as resonance peak fold-over \cite{Buchanan2007,Guslienko2010,Drews2012} and  core polarity switching \cite{Lee2008,Pigeau2010a,Pigeau2010}. The latter occurs when the core reaches a material-dependent critical velocity\cite{Lee2008}, $\vc$. Core switching (as well as core expulsion\cite{Jenkins2016}) open up attractive routes for controlling core polarity, critical  for data storage\cite{Pigeau2010}, field sensing\cite{Fried2016} and electronic oscillator tuning \cite{deLoubens2009,Dussaux2010}. 

Core polarity switching driven by dynamic phenomena has been probed predominantly using high resolution magnetic imaging techniques capable of resolving the  nano-scale  core's magnetization orientation \cite{Yamada2007,Kammerer2011,Yamada2014}. Magnetic resonance force microscopy has also been employed\cite{Pigeau2010a} and more recently, switching was detected using bench-top magneto-optics \cite{Dieterle2016}. It has been evidenced magnetoresistively in double vortex nano-pillars\cite{Locatelli2013a} and using a magnetic tunnel junction fabricated above the central portion of a magnetic disk  \cite{Nakano2010}, a measurement geometry which enabled real-time probing of the magnetization in the central part of the disk. 

A large number of measurements of steady state vortex gyration (as well as anti-vortex dynamics\cite{Goto2016}) have however  exploited simpler device geometries where laterally injected currents are used to  drive and probe the gyrotropic resonance\cite{Kasai2006,Bedau2007,Kim2010b,Sugimoto2011}. In these measurements, current-driven gyrotropic motion of the core leads to oscillations in the device resistance (via anisotropic magnetoresistance, AMR) which can mix with the input r.f.~current to generate a measurable, rectified d.c.~voltage. 

In this paper we show that this rectified signal is lost or strongly reduced when core polarity switching occurs, offering a simple, all-electrical method to probe core switching in single disks. Using this technique, we demonstrate how static applied fields can be used to control the range of frequencies over which core switching occurs by moving the core into a region of stronger confinement where the core is impeded from reaching $\vc$. We  observe signatures of peak fold-over and resonance downshifting  when increasing the excitation power, demonstrating the use of this simple detection scheme to probe regimes of highly non linear magnetization dynamics, which otherwise remain complex to study.

\begin{figure}[htbp]
	\includegraphics[width=8.4cm]{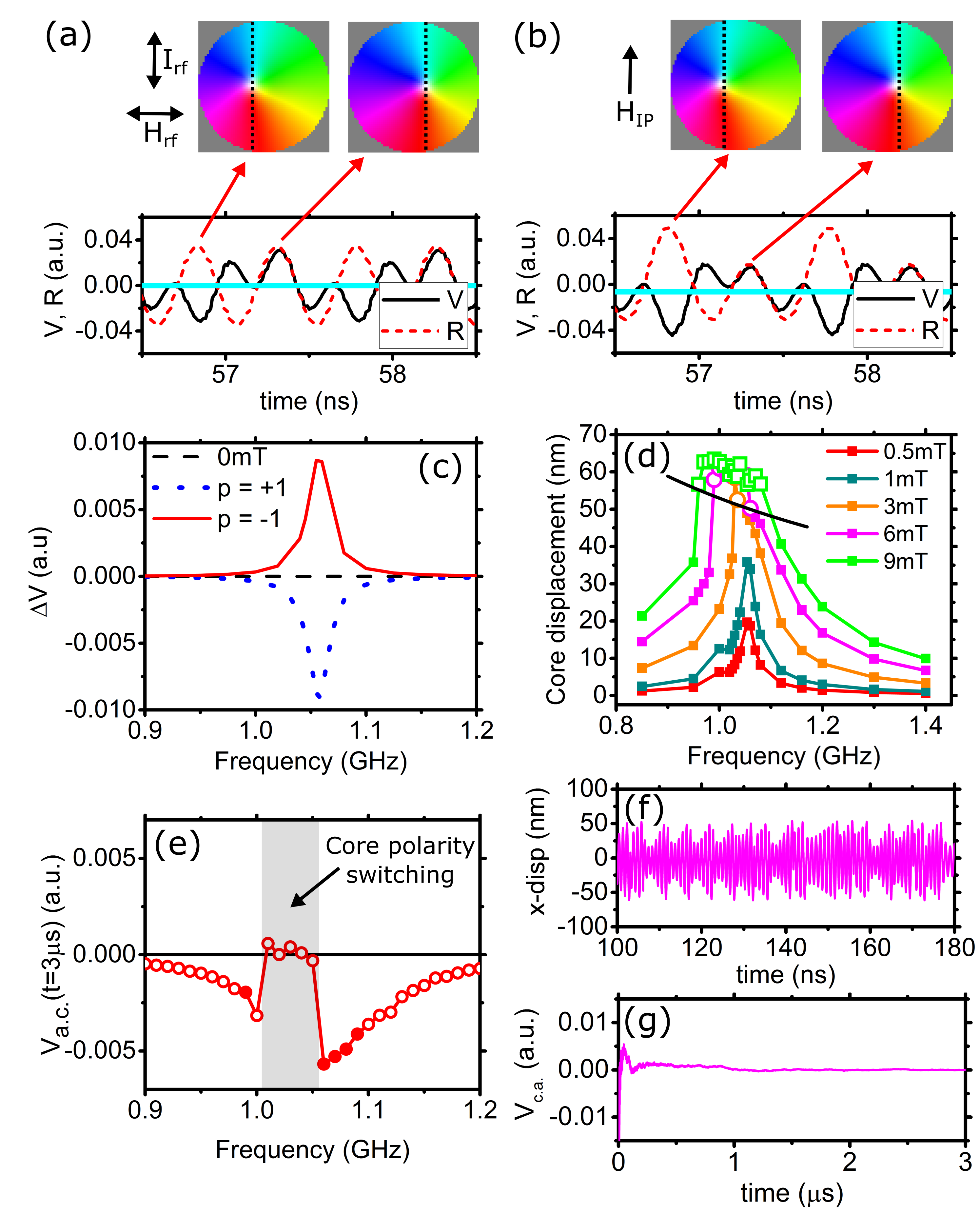}
	\caption{Dynamic resistance change and voltage due to core oscillations for $p=+1$ and $\hip =0$ (a) and $\mu_0\hip =+5$ mT (b). $f=1.05$ GHz and $\hrf =0.5$ mT. Images show  snapshots of the core positions with the solid dashed lines being reference positions of the left and right displaced cores for $\hip =0$. Solid light blue lines show the rectified (averaged) d.c.~voltage. (c) Resultant voltage peaks for $\mu_0\hip =+5$ mT ($p=+1$) and $\mu_0\hip =+5$ mT (for $p=\pm 1$) and $\mu_0\hrf =0.5$ mT obtained by averaging the time dependent $\Delta V$  over 50 oscillations of the core gyration period, commencing at $t=100$ ns  so as to use steady state dynamics.   (d) Core displacement versus $f$ and $\hrf$. The solid black line shows the $\fg$ obtained for a sinc pulse excitation. (e) Time averaged rectified signal (3 $\mu$s of averaging) versus frequency for  $\mu_0\hip =5$ mT and $\hrf = 6$ mT. (f) Core position versus time for a vortex undergoing core polarity switching. Switching events are identified from a reduced displacement. (g) Cumulative time average of the rectified voltage. }
		\label{fsim}
\end{figure}

We first simulate core gyration  in a 30 nm thick, 192 nm wide NiFe-like disk using  MuMax3\cite{Vansteenkiste2014} with  saturation magnetization $M_{\mathrm{S}}=800$ kA/m, exchange stiffness  $A_{\mathrm{ex}}=13$ pJ/m, nil intrinsic anisotropy and gyromagnetic ratio $\gamma=1.7595\times 10^{11}$ rad.(T.s)$^{-1}$. The cell size was $3\times 3 \times 3.75$ nm$^3$ ($64\times 64 \times 8$ cells). We initialize the system with a vortex-like state, optionally apply an in-plane static magnetic field, $H_{IP}$, to displace the core and then let the magnetization configuration relax. We then apply a sinusoidal in-plane excitation field, $H_{rf}=h_{rf}\sin (2 \pi f t)$, at frequencies, $f$, in the neighborhood of $\fg$. In the second part of the paper we will present measurements of core dynamics in wider disks where the 30 nm thick NiFe layer is covered by a non-magnetic capping layer of thickness $\approx 40$ nm). Currents, $I_{rf}=I_0\sin (2 \pi f t)$, flowing through the disk will generate an in-plane field in the lower NiFe layer which will be the dominant driver of the gyrotropic mode. We approximate this field  by $H_{rf}$ in the simulation which enables us to replicate many of the observed experimental features. The r.f.~field is transverse to the r.f.~current flow (consistent with an Oersted field), the latter being parallel to $\hip$ [Figs.~\ref{fsim}(a,b)]. 

AMR  depends on the angle between the \textit{axes} along which the current flow and magnetization are aligned.   During a single orbit of the core, the resistance  will increase twice since  left- and right-displaced cores both increase the disk resistance, $R$ [Fig.~\ref{fsim}(a)]. For symmetric electrode placement on a circular disk the change in $R$, $\Delta R$, is given by\cite{Goto2011,Goto2011a} $\Delta R\propto x^2-y^2$  where $(x,y)$ is the time dependent core position measured with respect to the disk's center. As such, a   core oscillating around the center of the disk at a frequency $f$   generates a resistance oscillation at $2f$. However,  there is no rectified voltage since the dynamic %$\Delta V\propto\Delta R \sin 2 \pi f t$ 
$\Delta V\propto\Delta R I_{rf}$ is symmetric around zero [light blue line in Fig.~\ref{fsim}(a)].  Although the $2f$ signal can be directly probed\cite{Sugimoto2014}, a finite rectified voltage can be achieved by laterally displacing the equilibrium core position using a static in-plane field \cite{Goto2011,Goto2011a}.  The changes in resistance for left and right displaced cores are then unequal [Fig.~\ref{fsim}(b)], the resulting $\Delta R$ having a component changing at $f$ which can mix with $I_{rf}$ (also at $f$) and generate a finite, rectified time averaged $\Delta V$ on resonance [Figs.~\ref{fsim}(b,c)]. Changing the core polarity  changes the direction of core gyration and thus the sign of the rectified voltage, generating a core-polarity-dependent voltage peak for $| \hip | >0$ at $\fg$  [Fig.~\ref{fsim}(c)].

A higher $\hrf$   generates much larger maximum core displacements during the oscillation [Fig.~\ref{fsim}(d)]. There is also a down shifting of the center of the resonance peak linked to fold-over. We  expect core switching\cite{Lee2008} to occur when the core reaches $\vc=333$ m/s ($=2 \pi\gamma 1.33\sqrt{A_{ex}}$; uncert.~$10.8$ \%). Thus, assuming a circular orbit, we expect the maximum  core displacement for which steady gyration can be driven to be $333/2\pi f$. Indeed, the core displacement data for which switching did and did not occur are shown respectively as open and closed symbols in Fig.~\ref{fsim}(d) with  the $333/2\pi f$ line shown in Fig.~\ref{fsim}(e) consistently separating the two sets of points.  

Since the sign of the rectified signal depends on the core polarity, a periodic switching of the polarity would  generate a null averaged $\Delta V$. Indeed, this is seen in simulation where a strongly reduced signal near resonance is observed [Fig.~\ref{fsim}(e)]. Note that polarity switching events are not perfectly regular  [Fig.~\ref{fsim}(f)]. Indeed, as seen previously \cite{PetitWatelot2012}, we found  a dependence of the switching  traces on small changes in the initial conditions of the simulation, consistent with chaotic dynamics. Irregular switching events were also  previously evidenced in experiment\cite{Nakano2010}. To accurately describe the time-averaged behavior of the turbulent $\Delta V$, we must look  at its cumulative average over long time scales: $V_{c.a.}(t)=\sum_{t'=0}^{t'=t}V(t')$ [Fig.~\ref{fsim}(g)]. Indeed, after $\approx$ 1 $\mu$s of turbulent dynamics, $V_{c.a}\rightarrow 0$ with the values in Fig.~\ref{fsim}(e) corresponding to\footnote{For the filled circles in Fig.~\ref{fsim}(e), there was a \textit{single} polarity switch during the early transient dynamics after the onset of the excitation field leading to a positive voltage which we set to negative for consistency with the other `no switching' data points which all generated a negative voltage.} $V_{c.a.}$ at $t=3$ $\mu$s. Notably, the signal is either strongly reduced or equal to zero where core polarity switching occurs. Thus,  even irregular switching will  lead to a quasi-null averaged rectified voltage at long enough time scales.

\begin{figure}[htbp]
	\includegraphics[width=8cm]{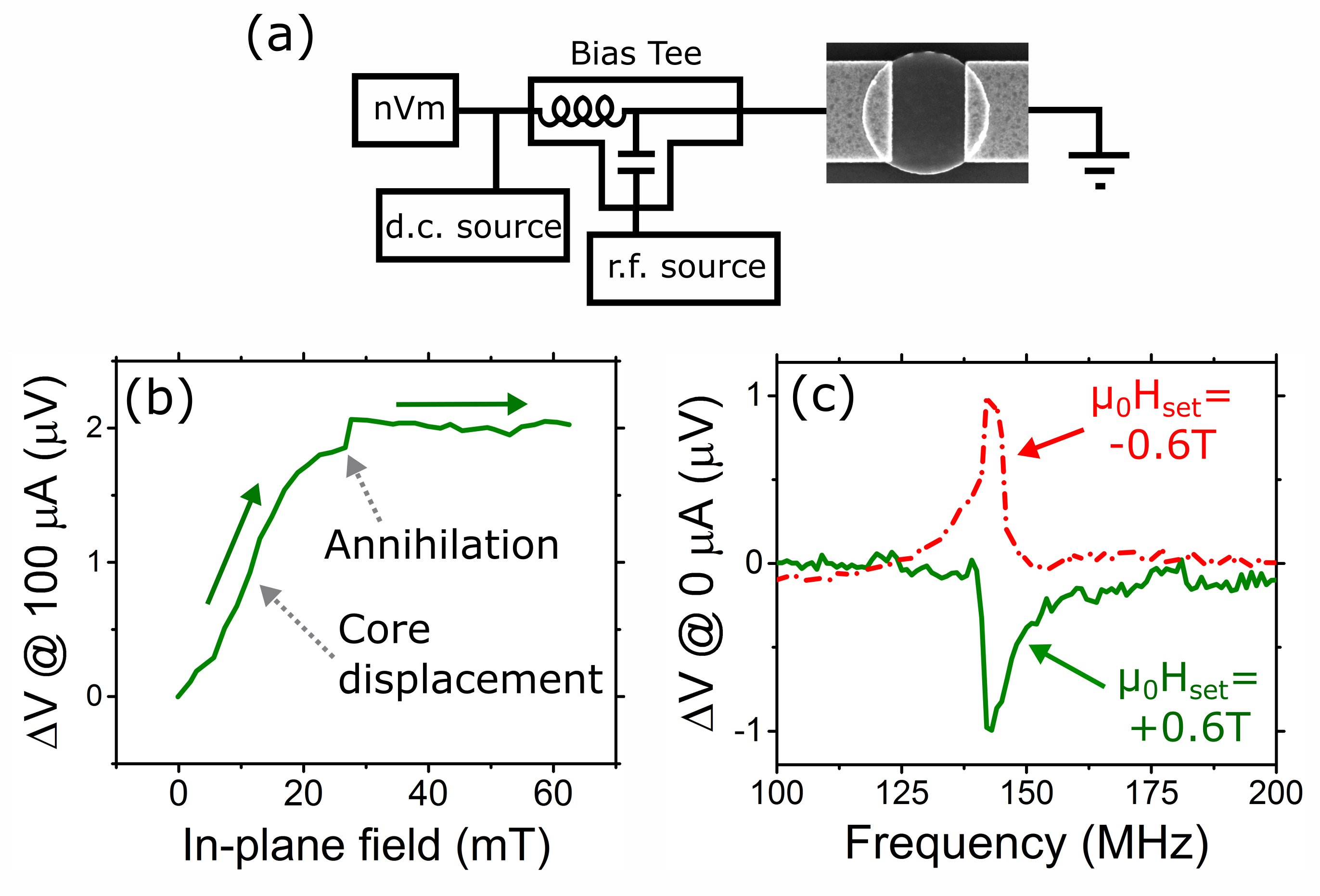}
	\caption{(a) Scanning electron micrograph of a 2 $\mu$m wide disk with lateral contacts together with the electrical circuit (`nVm'=nanovoltmeter). (b) Change in voltage across the measured 2 $\mu$m device  under a d.c.~current of 100 $\mu$A while increasing the in-plane field. The voltage change ($\approx$20 m$\Omega$) is characteristic of core displacement and vortex annihilation. (c) Low power rectification peaks observed at an in-plane field of $\approx 5 $ mT after subjecting the disk to $H_{set}$ fields at 45$^{\circ}$ to the sample plane.}
		\label{fexp}
\end{figure}

We now use  rectification to evidence core switching in 2 $\mu$m wide //NiFe(30 nm)/Au(10 nm)/Ti($\approx$ 30 nm) disks [Fig.~\ref{fexp}(a)]. They were fabricated from a continuous //NiFe(30nm)/Au(10 nm) layer via Argon ion milling using a Ti disk as a milling mask (defined via electron beam lithography and lift-off). Auger spectroscopy revealed that approximately 30 nm of Ti remained after etching, leading to the tri-layer structure referenced above.  Lateral contacts (//Au/Ti) were defined using electron beam lithography and lift-off. The device was wire bonded to a sample mount which could rotate between the poles of an electromagnet whose field was monitored using a Hall probe located near the mount. The rotation was such that the field could be applied along or out of the plane of the disk. Magnetoresistance measurements exploiting AMR  reveal core displacement followed by vortex annihilation as $\hip$ is increased [Fig.~\ref{fexp}(b)]. A bias tee enables the injection of an  r.f.~signal whose frequency is stepped during rectification measurements while measuring the d.c.~voltage across the device using a nanovoltmeter (200 ms integration time with injected d.c.~current $=0$ unless otherwise noted)

To test for core polarity dependence of the rectified voltage, we subjected  the device to a  field of $\mu_0 \hset =\pm 0.6$ T applied at an angle of 45$^{\circ}$ to the sample plane. After returning to 0$^{\circ}$, under a small field of $\mu_0 H_{IP}\approx 5$ mT  we obtained different polarity signals depending on the sign of $\hset$, consistent with a $\hset$-induced setting of the polarity [Fig.~\ref{fexp}(c)].  These curves were obtained under a low r.f.~power of -15 dBm. Experiments will be presented by referring to injected r.f.~powers in dBm with a comparison to field given at the end of the paper.

\begin{figure}[htbp]
	\includegraphics[width=8.4cm]{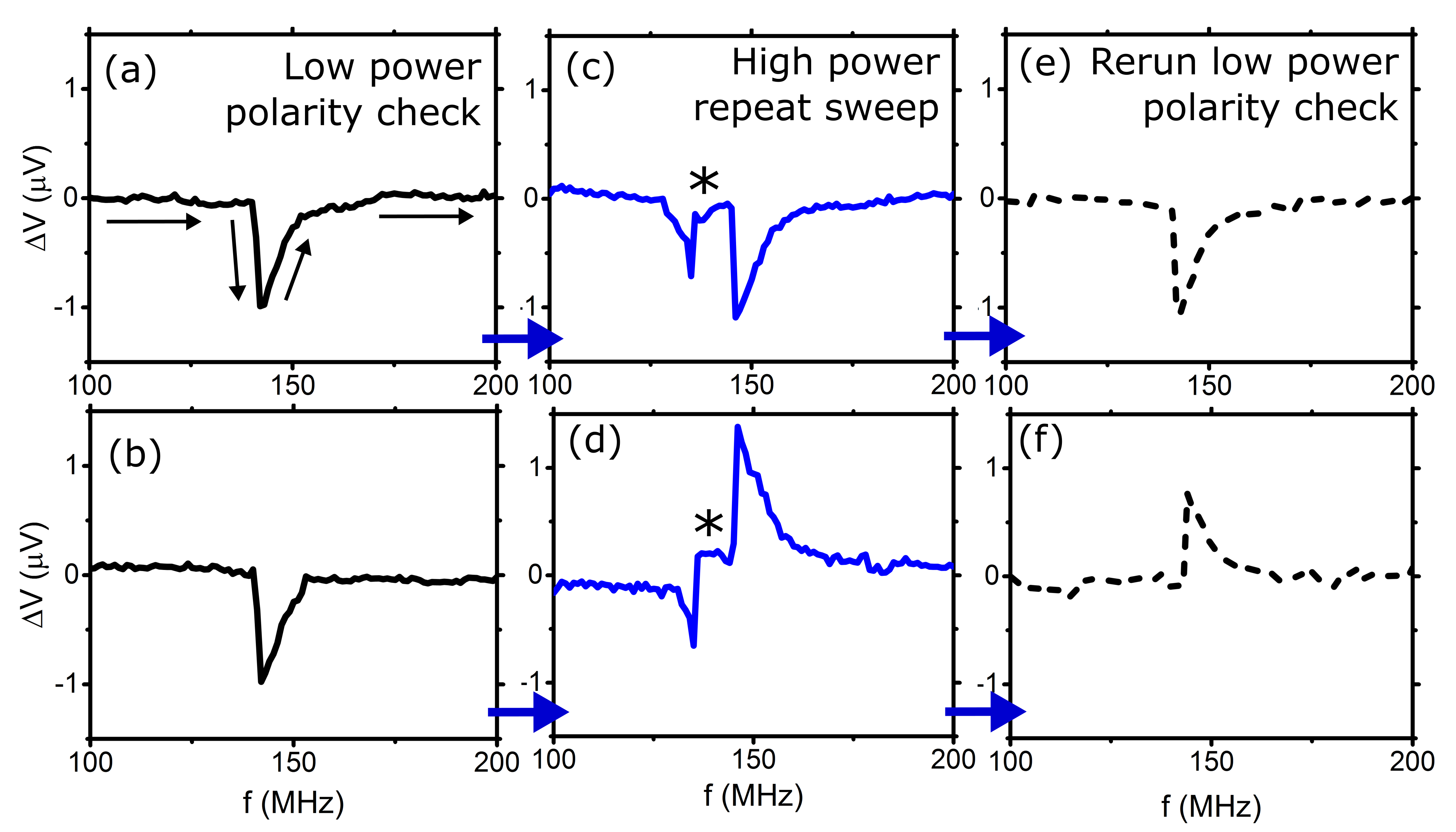}
	\caption{
	(a,b) Low power (-15 dBm) frequency sweep after setting a $p=-1$. (c,d) High power sweeps (-8 dBm) exhibiting signal loss (*). In (c), the polarity is the same after switching and in (d) the polarity has changed. (e,f) show low power sweeps confirming the post-signal-loss polarity.
	%2 $\mu$m disk P6634F_Apads
	}
		\label{fswitch}
\end{figure}

After setting and confirming a positive core polarity [Figs.~\ref{fswitch}(a,b)],  we   repeated the same frequency sweeps at a higher power [Figs.~\ref{fswitch}(c,d)]. Note that frequency sweeps were always from low frequency to high frequency [as per the black arrows in Fig.~\ref{fswitch}(a)]. As the frequency is increased for the high power rectification traces, we see the onset of a broad, downshifted negative voltage peak (consistent with the initial positive core polarity) which is followed by a region of strongly reduced $\Delta V$ (marked by a *) . After passing through the signal loss region, a finite $\Delta V$ is once again observed, consistent with a steady motion of the core generating the remainder of the rectification peak. The trace form is consistent with that simulated in Fig.~\ref{fsim}(e) where signal loss was due to a quasi-periodic core polarity switching. The second part of the rectification peak could have either polarity, in some cases reversing its sign, consistent with steady motion of a core with the opposite polarity to that prepared at the start of the experiment  [e.g. Fig.~\ref{fswitch}(d)]. The final (post-switching) core polarity could be confirmed using a second low power sweep. See Fig.~\ref{fswitch}(e) for an unchanged polarity and Fig.~\ref{fswitch}(f) for a switched polarity, the latter providing an alternative confirmation of polarity switching occurring as we sweep past the resonant frequency\footnote{We note that if the core was being expelled from the disk (as seen in [\onlinecite{Jenkins2016}] and here in  simulations of shifted cores in the 192 nm disk) then this would translate into a change in the d.c.~resistance of the disk on the order of 2 $\mu$V as per Fig.~\ref{fexp}(b). This would only be seen for a finite d.c. current however and rectification measurements carried out for a 100 $\mu$A current however did not lead to any measurable changes in $\Delta V$ in the switching region, thus ruling out expulsion.}. As simulated in Fig.~\ref{fsim}(e), the low frequency side of the observed peak is lower in magnitude and the peak signal starts growing at a lower frequency. This is consistent with a non-linear fold-over effect which leads to a downshifting of the resonance frequency (compared to that obtaiend at low power) and a  sharper onset of the resonance peak (and thus switching events) at the lower frequency side of the peak.

\begin{figure}[htbp]
	\includegraphics[width=8.4cm]{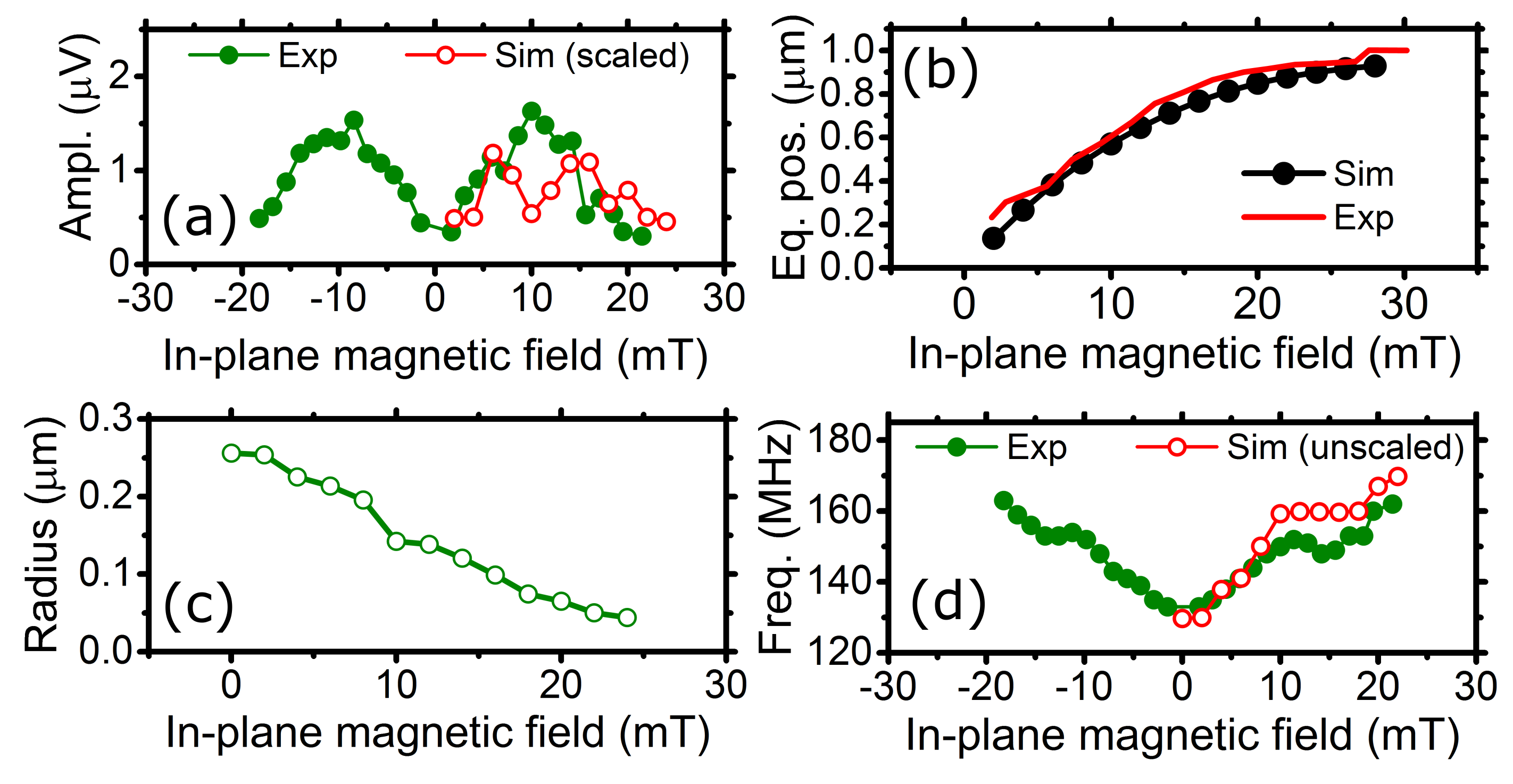}
	\caption{(a) Measured voltage peak amplitude versus $\hip$ together with the simulated amplitude (scaled to be comparable to the experimental data at its maximum). (b) Equilibrium core position simulated in a 2 $\mu$m disk versus $\hip$. The solid line is the core position extracted from the data in Fig.~\ref{fexp}(b) as per the text. (c) Core-radius defined as (max(x)-min(x))/2 during steady state core motion where x is the core position. (d) Frequency versus $\hip$ for -15 dBm and reproduced via simulation.}
		\label{famp}
\end{figure}

We now briefly return to lower power excitations before looking at how static in-plane fields affect core polarity switching. We will compare experimental data with simulation results obtained for a 2 $\mu$m wide, 30 nm thick disk on a large $768\times 768 \times 1$ mesh (i.e.~with a \textit{single} discretization cell\footnote{Tests in the 192 nm $\times$ 30 nm disk demonstrated that while switching was observed when using 8 cells in the $z-$direction, it was no longer seen under the same conditions for a single cell.  Thus we look only at low power excitation ($\hrf=0.2$ mT) for the 2 $\mu$m disk.} in the $z-$direction to make the simulation more tractable; lateral cell size was $2.6\times 2.6$ nm$^2$). 
In Fig.~\ref{famp}(a) we show the experimental amplitude of the rectified low power signal due to core gyration with no switching for a range of $\hip$ values. The initial increase when moving away from $\hip =0$ arises due to the increasingly offset equilibrium core position [Fig.~\ref{famp}(b)] which generates more asymmetric variations in the resistance, leading to a larger $\Delta V$ [as per  Figs.~\ref{fsim}(a,b)]. It is important to note though that the core displacement is clearly non-linear in $\hip$, consistent with an anharmonic confining potential\cite{Sukhostavets2013} which results in the core stiffness increasing towards disk edge\cite{Sukhostavets2013,Fried2016}. There is notably good agreement between the simulated and experimental core positions with the latter determined from the pre-annihilation data in Fig.~\ref{fexp}(b) using $x = \sqrt{\Delta R/\Delta R_{max}}$ (in $\mu$m, supporting the use of $\Delta R\propto x^2-y^2$). =
A consequence of the increased edge stiffness is a reduced gyration radius for diaplced cores [Fig.~\ref{famp}(c)], consistent with the smaller $\Delta V$  at large $\hip$. The simulated signal amplitude [arbitrarily scaled for comparison to experiment in Fig.~\ref{famp}(a)] reproduces the amplitude increase for increasing low $\hip$ as well as the amplitude decrease at high $\hip$. The simulation however predicts a drop off at intermediate $\hip$ (where the oscillating $\Delta V$ is a deformed sinusoid) which was not observed in experiment.  Another consequence of the anharmonicity is that the frequency  is $\hip$-dependent\cite{Sukhostavets2013}, as shown experimentally in Fig.~\ref{famp}(d) and reproduced via `ringdown'\cite{Fried2016} simulations to within 7\%.

\begin{figure}[htbp]
	\includegraphics[width=8.5cm]{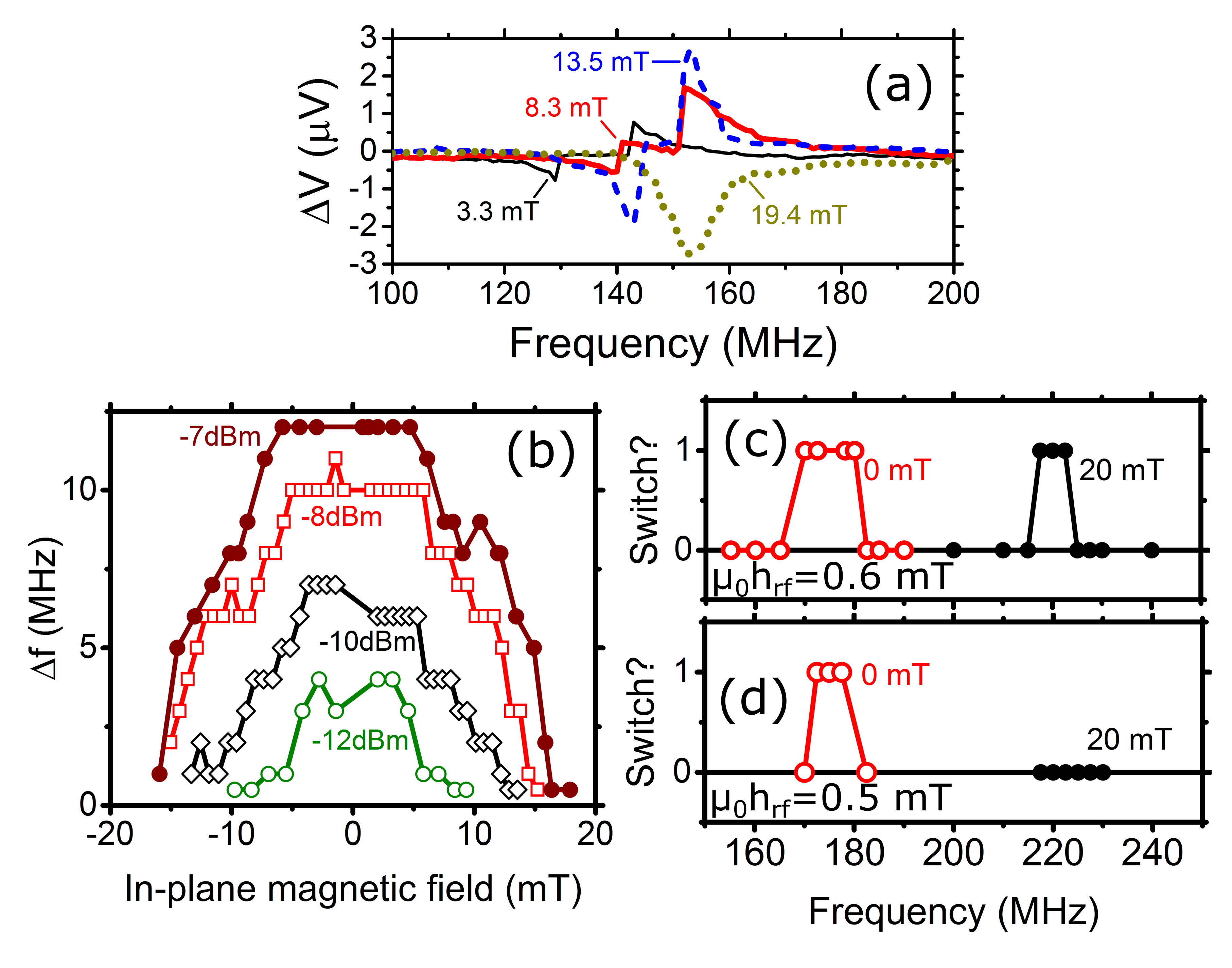}
	\caption{(a) High power rectification curves (-7 dBm) versus field showing how the signal loss region reduces in size and eventually disappears for increasing $\hip$. (b) Range of frequencies over which signal loss occurs versus $\hip$ for various r.f.~powers. (c,d) This simulation data shows if a core in a 1.4 $\mu$m disk switched (``1'') or did not switch (``0'')  under $\hip =$ 0, 20 mT  for $\hrf=0.6$ mT (c) and $\hrf=0.5$ mT (d). }
		\label{fdelta}
\end{figure}

 Fig.~\ref{fdelta}(a) shows experimental rectification traces in which the signal loss region is also seen to reduce (and eventually vanish) with increasing $\hip$.  Note that the resonance also shifts rightward, consistent with the anharmonic confinement. Indeed, for all powers, the range of frequencies, $\Delta f$, over which signal loss, and thus switching, occur reduce with $\hip$ as shown directly in Fig.~\ref{fdelta}(b). Note also that, as is expected from Fig.~\ref{fsim}(d) and seen previously\cite{Dieterle2016},  $\Delta f$ can be increased by  increasing the  power since $\vc$ will be reached for a larger range of frequencies.
The impeded switching at the core edge (i.e.~for increased $\hip$) is consistent with an increased confinement which would be expected to impede the core from reaching $\vc$. To confirm this, core switching was simulated in a similarly sized disk (1.4 $\mu$m) where a full $z-$discretization (8 cells) was computationally tractable\footnote{Due to a reduced number of lateral cells ($512\times 512$)}.  Applying a finite $\hip$ shifts the vortex, increasing $\fg$ and thus the frequencies at which switching occurs [Fig.~\ref{fdelta}(c)]. For $\hrf= 0.6$ mT, the increased $\hip$ however also reduces the range of frequencies over which switching occurs  [Fig.~\ref{fdelta}(c)] with no switching observed for the shifted core under $\hrf= 0.5$ mT [Fig.~\ref{fdelta}(d)].  The absence of switching is indeed accompanied by a reduction in the average velocity during the non-circular orbit of the shifted vortex: $\bar v\approx 285$ m/s (within 14\% of $\vc$) for the centered, switching core and  $\bar v\approx 264$ m/s for the shifted, non-switching core.  Note that in simulations of the 1.4 $\mu$m disk, core polarity switching for non-displaced vortices was observed first at a $\hrf$ of 0.5 mT. This is encouragingly of the same order of magnitude as the field estimated to act at the bottom of the NiFe layer in the disk centre  ($\sim 0.27$ mT) during the lowest power (-12 dBm)   switching observed experimentally in a 1.4 $\mu$m disk (assuming uniform current flow).

In conclusion, we have presented an all-electrical measurement of core polarity switching via magnetoresistive rectification in magnetic disks. Turbulent core switching together with a core polarity dependent signal leads to loss of the rectified voltage upon switching. We demonstrate that core switching is impeded when the core approaches the disk edge, consistent with an increased confinement  which impedes the core from resonantly reaching its critical velocity.

This research was supported by the Australian Research Council's Discovery Early Career Researcher Award scheme (DE120100155), a research grant from the United States Air Force (Asian Office of Aerospace Research and Development, AOARD) and the University of Western Australia's Early Career Researcher Fellowship Support scheme. The authors acknowledge resources provided by the Pawsey Supercomputing  Centre with funding from the Australian Government and the Government of    Western Australia as well as the facilities, and the scientific and technical assistance of the Australian Microscopy \& Microanalysis Research Facility at the Centre for Microscopy, Characterisation \& Analysis, The University of Western Australia, a facility funded by the University, State and Commonwealth Governments. We thank F.~Wyczisk, D.~Turner, G.~Light, J.~Moore, D.~McPhee and M.~Frankenberger for their assistance and J.-V.~Kim for useful discussions.

\end{document}